\documentclass{ws-ijgmmp}
\newcommand{\be}{\begin{equation}}
\newcommand{\ee}{\end{equation}}
\newcommand{\bea}{\begin{eqnarray}}
\newcommand{\eea}{\end{eqnarray}}
\def\vh{\varphi}
\begin{document}

\markboth{B.M. Barbashov, V.N. Pervushin, A.F. Zakharov, and V.A.
Zinchuk} {The Hamiltonian approach to General Relativity and CMB
primordial spectrum}

%
\catchline{}{}{}{}{}
%

\title{THE HAMILTONIAN
APPROACH TO GENERAL RELATIVITY AND CMB PRIMORDIAL
SPECTRUM
 }

\author{B.M. BARBASHOV 
}

\address{Joint Institute for Nuclear Research, Dubna, 141980, Moscow region, Russia\\
\,
}

\author{V.N. PERVUSHIN}

\address{Joint Institute for Nuclear Research, Dubna, 141980, Moscow region, Russia\\
pervush@theor.jinr.ru }

\author{A.F. ZAKHAROV}

\address{Institute of Theoretical and Experimental Physics,  117259, Moscow, Russia\\
zakharov@itep.ru }

\author{V.A. ZINCHUK}

\address{Joint Institute for Nuclear Research, Dubna, 141980,
Moscow region, Russia
 }

\maketitle

\begin{history}
\received{(01.06. 2006)} 
\end{history}

\begin{abstract}

Approaches to solutions of problems of the energy, time,
Hamiltonian operator quantization of the General Relativity, the
creation of the Universe from
 vacuum are considered in the frame of reference associated with
  the CMB radiation  in order to
 describe parameters of this radiation in terms of the parameters
 of the Standard Model of elementary particles.
\end{abstract}


\section{Introduction}

 The measurements of the
 Cosmic Microwave Background (CMB) radiation
  temperature  \cite{WMAP} revealed its dipole component
 testifying that an Earth observer (i.e. his ``Hubble''
  Telescope)  moves to Leo with a velocity about 400 km/c.
   This velocity can be treated as a parameter of the
   Lorentz transformation between a rest frame of an  Earth observer
 and the comoving frame of the
 Universe
   distinguished by the
  unit  time-like vector ($l_\mu=(1,0,0,0)$).

  The revelation
  of the comoving frame of our Universe allows us to
  seek
  explanations of  cosmological problems, or part of them,
   with the help of the ordinary Laplace-type questions:
  What are  primordial values
  of the cosmological scale factor  and field variables
  in the comoving frame?
  What are the units of measurement of the
   initial data
  which can give us the simplest fit of all observational data?

 In order to  describe the Universe creation and its evolution
  in the comoving reference frame,
 we apply the Hamiltonian  approach to  General Relativity (GR) and Standard Model (SM)
  of  elementary particles
  developed  in the case of QED by Dirac \cite{Dir} who
   distinguished the comoving frame by the unit  time-like vector
   $l_\mu$
  and separated  vector field components $A_\mu$
  into time-like $A_0=l_\mu A_\mu$ and
  the space-like ones $A_j$. The latter are split
  on two transverse degrees of  freedom\footnote{We shall use symbols
  of the Laplacian $\triangle=\partial_k^2$ and d'Alambertian
  $\Box=\partial_0^2-\triangle$ determining a difference
  between the potential  $\triangle A_0=-j_0$ and
  degrees of freedom  $\Box A_k=-j_k$.}  $A^{(\rm T)}_i=
  (\delta_{ij}-\partial_i\frac{1}{\triangle}\partial_j)A_j$
   (called the ``photon'')
   and the longitudinal part $A^{(||)}_i=
   \partial_i\frac{1}{\triangle}\partial_jA_j$
   which together with the time-like component $A_0$ form
   the gauge-invariant Coulomb  potential $A_0^{(\rm
   T)}=A_0-\partial_0\frac{1}{\triangle}\partial_jA_j
   $.  Dirac \cite{Dir} called these gauge-invariant
   functionals $A^{(\rm T)}_0,A^{(\rm T)}_k,\Psi^{(\rm T)}
   =\exp\{ie\frac{1}{\triangle}\partial_jA_j\}\Psi$
   as the ``dressed'' fields (where
   $e$ is a coupling constant).
   We can {\it quantize} only two gauge-invariant transversal components,
    while
   the {\it instantaneous} Coulomb
 potential $A_0^{(\rm T)}$
  forms {\it instantaneous} atoms and
 molecules\footnote{Recall that the  Faddeev-Popov heuristic
  quantization  \cite{fp1} in the frame free ``Lorentz gauge
 formulation'' leads to
  photon propagators  having only the light cone singularities.
 In fact, Faddeev \cite{f1} proved the
 equivalence of
  this frame free gauge formulation with the Dirac
 {\it operator quantization}  \cite{Dir}
  only for the elementary particle scattering amplitudes
   \cite{pvn3}, which do not depend on initial data.}.

In the context of a consistent description of
   bound states and collective evolution
 of the type of cosmological expansion,
 one can ask what  the {\it instantaneous} Newton potential and
  {\it the  operator
quantization} of GR are, if  the cosmological evolution is
considered as one of dynamic variables that can be quantized?

\section{Hamiltonian Approach to GR in Finite Space-Time}


 Einstein's GR \cite{einsh} is associated with Hilbert's action  \cite{H}
 \be
 \label{gr}
 S_{\large \rm GR}[\vh_0|F]=\int
d^4x\sqrt{-g}\left[-\frac{\vh_0^2}{6}R(g)
 +{\cal L}_{(\rm M)}\right],
 \ee
where $\vh_0=M_{\rm
 Planck}\sqrt{3/{8\pi}}$ is the Planck
mass parameter. Recall that Hilbert  \cite{H} formulated GR so
that Einstein's generalization of the Lorentz frame group
$x^\mu\to \widetilde{x}^\mu
 =\widetilde{x}^\mu(x^\mu)$ becomes a gauge group.
 There is the principal difference between
 the frame transformations and the gauge ones.
 Parameters of  frame transformations (of type of initial data)
 are treated as measurable quantities,
  whereas parameters of the gauge  transformations
  are not measured.  Gauge symmetries lead to constraints
  decreasing number of degrees of freedom.

  In the context of the problem of
  the initial data one needs to separate the frame transformations
  (here the Lorentz ones)
  from the gauge transformations (here the general coordinate
 ones).  Just this separation is the main difference
  of the Hamiltonian approach to
 GR considered here in finite space-time \cite{pvng8a,pvng8b}
  from the Dirac -- ADM one \cite{dir}.
 This separation
 can be fulfilled by using  the gauge-invariant components
  of Fock's symplex $\omega_{(\alpha)}$ defined as
$
 ds^2=g_{\mu\nu}dx^\mu dx^\nu=\omega_{(0)}\omega_{(0)}-
 \omega_{(b)}\omega_{(b)};~~\omega_{(\alpha)}=e_{(\alpha)\nu} dx^\nu
 $
 where
 $e_{(\alpha)\nu}$ are
 the Fock tetrad the components of which   are marked by
 the
 general coordinate index without a bracket and
 the Lorentz index in  brackets $(\alpha)$ \cite{fock29}.

 The choice of   the time axis $l_{(\mu)}=(1,0,0,0)$ as the
 CMB comoving frame allows us to construct
 an irreducible representation of the Poincare group by
  decomposition of Fock's  vector simplex field $\omega_{(\alpha)}$
  in accordance with the  definition of the Dirac--ADM Hamiltonian
  approach to GR \cite{dir}
  \be\label{om}
  \omega_{(0)}=\psi^6N_{\rm d}dx^0\equiv \psi^2\omega^{(L)}_{(0)},~~~
  \omega_{(b)}=\psi^2 {\bf e}_{(b)i}(dx^i+N^i dx^0)\equiv\psi^2
  \omega^{(L)}_{(b)},
  \ee
 where $N^i$ is shift vector,
 $N_{\rm d}$ is Dirac's lapse function, $\psi$ is the spatial metric
 determinant, ${\bf e}_{(b)j}$ is a triad with
 the unit determinant $|{\bf e}|=1$,
 and $\omega^{(L)}_{(0)},\omega^{(L)}_{(b)}$ are
 the scale-invariant Lichnerowicz simplex \cite{Y}
 forming the scale-invariant volume
 $
 dV_0 \equiv
 \omega^{(L)}_{(1)}\wedge
 \omega^{(L)}_{(2)}\wedge\omega^{(L)}_{(3)}=d^3x^{(L)}
 $
 that coincides with the spatial coordinate volume.


 It is well known \cite{6} that
 the Hamiltonian approach to GR is invariant
 with respect to
 the general spatial coordinate transformations $x^0 \to
 \widetilde{x}^0=\widetilde{x}^0(x^0)$  and
 $x^j \to \widetilde{x}^j=\widetilde{x}^j(x^0,x^j)$.

 The gauge invariance $x^j \to \widetilde{x}^j=\widetilde{x}^j(x^0,x^j)$
 allows us to remove
 longitudinal components of tensor
 triads  ${\bf e}_{(b)j}$ and keep
 two transversal gravitons distinguished by the constraint
 $
\partial_i{\bf e}^{(\rm T)i}_{(b)}\simeq 0
 $ in complete analogy with
  the Dirac construction of
 QED with the one-to-one
 correspondence $[ A^{(\rm T)}_0,A^{(\rm T)}_k]~\to~[N^{(\rm T)}_{(b)},
 e^{(\rm T)}_{(b)k}]$.
 This means that the  spatial coordinates
 and the Lichnerowicz finite volume $V_0=\int d^3x^{(L)}=\int d^3x$ can be identified
 with  gauge-invariant observables.

 The  invariance of GR in the  finite volume $V_0$ (considered in the modern
 cosmology for description of the Universe evolution)
 with respect to the reparametrizations of the
 coordinate evolution parameter: $x^0 \to
 \widetilde{x}^0=\widetilde{x}^0(x^0)$ \cite{6}
 allows us to convert one of variables into the time-like evolution
 parameter in the comoving frame.
    This means that
 the coordinate evolution parameter $x^0$ is not
 observable. Wheeler and DeWitt \cite{WDW} proposed considering the
 reparametrization invariance
 in GR by analogy with   Special Relativity
 (SR), where  the role of a timelike variable is played by
   one of the dynamic
 variables  $X_0$ in the World space of events $[X_0|X_k]$.

 Wheeler and DeWitt \cite{WDW} proposed considering the
 reparametrization invariance
 in GR in a similar manner, i.e. they proposed to generalize
 the construction of  representations
 of the Poincare group
 to the field space of events in GR, where
 the role of
 reparametrization-invariant evolution parameter
   (i.e. a time-like variable in the field space of events) is
   played by
  a  cosmological scale factor $a(x^0)$. This factor is
  separated by the scale transformation of all fields with a conformal
  weight $n$ including
 the metric components
 $
 F=a^n(x^0) \widetilde{F}^{(n)},~~
 g_{\mu\nu}=a^2(x^0)\widetilde{g}_{\mu\nu}
 $.
 This transformation keeps the momentum constraint
 $T^k_0=\widetilde{T}^k_0=0$, so that
 the cosmological scale factor $a(x^0)$ can be
 considered as the zero mode solution
 of the momentum constraint. In order to conserve the number of variables in
   GR, the logarithm of
 the cosmological scale factor is identified with  Lichnerowicz spatial
 averaging  the  spatial determinant logarithm
 $ 
  \log a = \langle \log \psi^2\rangle\equiv \frac{1}{V_0}\int d^3x\log
  \psi^2.
 $ 
 This  finite volume generalization of the
 Dirac--ADM approach \cite{dir} was called
 the ``Hamiltonian cosmological perturbation
 theory''  \cite{pvng8b} \footnote{
 The separation of the cosmological scale factor is well-known
 as the ``cosmological perturbation theory''
 (where $\widetilde{\psi}=1-\Psi/2,\widetilde{\psi}^6\widetilde{N}_{\rm d}=1+\Phi$)
  proposed by Lifshits
 \cite{lif} in 1946 and applied now for analysis of
 observational data in modern astrophysics and cosmology
  (see  \cite{bard}).
 However, in this theory, the  scale factor is an additional variable without
 any constraint for the deviation $\Psi$, so that
 $\int d^3x \Psi\not = 0$, and
 there are two zero Fourier harmonics
 of the determinant logarithm
 instead of one. This doubling
 does not allow to express the velocities of
 both variables $\log a$ and $\int d^3x \Psi$ through
 their momenta and to construct the Hamiltonian approach to GR
 \cite{pvng8a,pvng8b}.}. This approach demonstrates that the naive
 perturbation theory $g_{\mu\nu}=g^{\rm
 Minkowski}_{\mu\nu}+h_{\mu\nu}$,
 where the coordinate evolution parameter $x^0$ is identified with
 measurable quantities, is incorrect because this theory
 contradicts to the gauge symmetry of GR.
 The definition of a measurable gauge-invariant coordinate
  evolution parameter in GR in
 finite volume will be considered below.


 A scale transformation
 $
 \sqrt{-g}R(g)=a^2\sqrt{-\widetilde{g}}R(\widetilde{g})-6a
 \partial_0\left[{\partial_0a}\sqrt{-\widetilde{g}}~\widetilde{g}^{00}\right]$
   converts action (\ref{gr}) into
 \be\label{1gr}
 S=\widetilde{S}-
 \int dx^0 \left(\frac{d\vh}{dx^0}\right)^2\int\limits_{V_0}
  {d^3x}{\widetilde{N}_{\rm d}}^{-1}\equiv  \int dx^0 L,
 \ee
 where $\widetilde{S}$
  is the action (\ref{gr})  in
 terms of metrics $\widetilde{g}$ and
 the running scale  of all masses
 $\vh(x^0)=\vh_0a(x^0)$. The variation of this action with respect
 to the new lapse function $\widetilde{N}_{\rm d}$ leads to a new energy constraint
 \be\label{nph}
  T^0_0=\widetilde{T}^0_0-
  \left(\frac{d\vh}{dx^0}\right)^2\frac{1}{\widetilde{N}_{\rm d}^2}=0
  ~~~~~~~~
  \left(\widetilde{T}^0_0=-
  \frac{\delta \widetilde{S}}{\delta \widetilde{N}_{\rm d}}\right).
 \ee
Spatial averaging the square root
 $\sqrt{\widetilde{T}^0_0}=\pm[d\vh/(\widetilde{N}_{\rm d}dx^0)]$
 over the Lichnerowicz volume $V_0=\int d^3x$
 gives  the Hubble-like relation
 \be\label{Lint}
 \zeta_{(\pm)}=\int dx^0\langle \widetilde{N}_{\rm d}^{-1}\rangle^{-1}=
  \pm\int^{\vh_0}_{\vh}
  d\vh/\langle{(\widetilde{T}_0^0})^{1/2}\rangle,
  \ee
   where
  $\langle F\rangle=V_0^{-1}\int d^3x F$
  and  $(d\zeta)^{-1}=\langle(dx^0\widetilde{N}_{\rm d})^{-1}\rangle$ is
  an
 inverse time-interval invariant with respect to
  time-coordinate transformations
  $x^0 \to \widetilde{x}^0=\widetilde{x}^0(x^0)$.
  We see that the  Hubble law in the exact
 GR  appears as  spatial averaging  the energy constraint (\ref{nph}).
 Thus, in the contrast with the generally accepted Lifshits
 theory \cite{lif} its Hamiltonian version \cite{pvng8b} distinguishes
 the time-coordinate $x^0$ as an object of reparametrizations
 from the reparametrization-invariant  time interval
 (\ref{Lint}).

 Just this distinction help us to separate
 the local part of the energy constraint (\ref{nph})
 and determine unambiguously
 the gauge-invariant
  Dirac lapse function
 \be\label{13ec}
 N_{\rm inv}={\langle(\widetilde{N}_{\rm d})^{-1} \rangle
 \widetilde{N}_{\rm d}}=
 {{\langle{(\widetilde{T}_0^0})^{1/2}\rangle}}
 (\widetilde{T}_0^0)^{-1/2}.
 \ee

The explicit dependence of $\widetilde{T}_0^0$ on
$\widetilde{\psi}$
  can be given in terms of the scale-invariant  Lichnerowicz
  variables  \cite{Y}
  $\omega^{(L)}_{(\mu)}=\widetilde{\psi}^{-2}\omega_{(\mu)},
  F^{(Ln)}=\widetilde{\psi}^{-n}\widetilde{F}^{(n)}$
 \be\label{t00}
 \widetilde{T}^0_0= \widetilde{\psi}^{7}\hat \triangle \widetilde{\psi}+
  \sum_I \widetilde{\psi}^Ia^{I/2-2}\tau_I, \ee
   where $\hat \triangle
 \widetilde{\psi}\equiv({4\varphi^2}/{3})\partial_{(b)}\partial_{(b)}\widetilde{\psi} $ is
 the Laplace operator and
  $\tau_I$ is partial energy density
  marked by the index $I$ running a set of values
   $I=0$ (stiff), 4 (radiation), 6 (mass), 8 (curvature), 12
   ($\Lambda$-term)
in accordance with a type of matter field contributions, and $a$
is the scale factor \cite{pvng8a}.

 The expression
 $(\widetilde{T}_0^0)^{1/2}$ is Hermitian
 if  a negative
 contribution of  the local determinant momentum
 \bea
\label{gauge}
 p_{\widetilde{\psi}}&=&\frac{\partial {\cal L}}{\partial (\partial_0\log\widetilde{\psi})}\equiv
 -\frac{4\vh^2}{3}\cdot\frac{\partial_l(\widetilde{\psi}^{6}N^l)-
 \partial_0(\widetilde{\psi}^{6})}{\widetilde{\psi}^{6}\widetilde{N_{\rm d}}},
 \eea
is removed  from the energy density (\ref{nph}) by the minimal
surface constraint \cite{dir}
 \be\label{hg} {p_{\widetilde{\psi}}}\simeq 0 ~~~~\Rightarrow ~~~~
\partial_j(\widetilde{\psi}^6{\cal N}^j)=(\widetilde{\psi}^6)'
~~~~~ ({\cal N}^j=N^j\langle \widetilde{N}_{\rm d}^{-1}\rangle).
 \ee
One can see that the scalar sector $N_{\rm
int},\widetilde{\psi},\partial_j [\widetilde{\psi}^6{\cal N}^j]$
is completely determined in terms of gauge-invariant quantities by
the equations  (\ref{13ec}), (\ref{hg}) and the equation of the
local part of the spatial determinant logarithm
 $\log\widetilde{\psi}^2\equiv\log \psi^2-\langle \log \psi^2\rangle$
 \be\label{lsf}
 \frac{\delta S}{\delta \log \widetilde{\psi}}\equiv
 -T_{\psi}+\langle T_{\psi}\rangle=0,~~~~~~~~
  \int d^3x\log\widetilde{\psi}^2\equiv 0,
 \ee
 where ${T}_{{\psi}}$  is given as
\be\label{tp}
 {T}_{{\psi}}|_{(p_{\widetilde{\psi}}=0)}=7N_{\rm inv}
 \widetilde{\psi}^{7}\hat \triangle \widetilde{\psi}+
  \widetilde{\psi}\hat \triangle[N_{\rm inv}\widetilde{\psi}^{7}]+
  \sum_I I\widetilde{\psi}^Ia^{I/2-2}\tau_I. \ee

These equations are in agreement with the Schwarzschild-type
solution for the potentials
$\triangle\widetilde{\psi}=0,\triangle[N_{\rm
inv}\widetilde{\psi}^{7}]=0$ in the empty space $\tau_I=0$, but
they strongly differ
  from the ``gauge-invariant'' version
\cite{bard} of the Lifshits perturbation
 theory   \cite{lif}.

 The Lifshits theory \cite{bard}
 does not take into account
  both the Dirac  constraint (\ref{hg}) removing
  the negative energy of the spatial determinant and
   the potential scalar perturbations formed by the
    determinant
   $\widetilde{\psi}=1+(\mu-\langle\mu\rangle) $ in (\ref{t00}), where
$\sum_I \widetilde{\psi}^Ia^{I/2-2}\tau_I=\sum_n
c_n(\mu-\langle\mu\rangle)^n\tau_{(n)}$,
 $\tau_{(n)}
 \equiv\sum_II^na^{\frac{I}{2}-2}\tau_{I}\equiv \langle\tau_{(n)}\rangle +
 \overline{\tau}_{(n)}$, and
 $\overline{\tau}_{(n)}={\tau}_{(n)}-\langle\tau_{(n)}\rangle$.
 The Hamiltonian cosmological perturbation theory \cite{pvng8a,pvng8b} leads
 to the scalar potentials
\bea\label{12-17}
 \widetilde{\psi}&=&1+\frac{1}{2}\int d^3y\left[D_{(+)}(x,y)
\overline{T}_{(+)}^{(\psi)}(y)+
 D_{(-)}(x,y) \overline{T}^{(\psi)}_{(-)}(y)\right],\\\label{12-18}
 N_{\rm inv}\widetilde{\psi}^7&=&1-\frac{1}{2}\int d^3y\left[D_{(+)}(x,y)
\overline{T}^{(N)}_{(+)}(y)+
 D_{(-)}(x,y) \overline{T}^{(N)}_{(-)}(y)\right],
  \eea
 where
 \be\label{1cur1}\overline{T}^{(\psi)}_{(\pm)}=\overline{\tau}_{(0)}\mp7\beta
  [7\overline{\tau}_{(0)}-\overline{\tau}_{(1)}],
 ~~~~~~~
 \overline{T}^{(N)}_{(\pm)}=[7\overline{\tau}_{(0)}-\overline{\tau}_{(1)}]
 \pm(14\beta)^{-1}\overline{\tau}_{(0)}
 \ee
 are the local currents, $D_{(\pm)}(x,y)$ are the Green functions satisfying
 the equations\footnote{In contrast
    to the Lifshits theory,
    the solutions (\ref{12-17}) and (\ref{12-18}) contain
    the nonzero shift-vector ${\cal N}^i$ of the coordinate origin
   with the spatial metric oscillations that lead  to a new mechanism of formation
  of the large-scale structure of the Universe \cite{pvng8a,pvng8b}.
 }
 \bea\label{2-19}
 [\pm \hat m^2_{(\pm)}-\hat \triangle
 ]D_{(\pm)}(x,y)=\delta^3(x-y)-{1}/{V_0},
 \eea
 where $\hat m^2_{(\pm)}= 14 (\beta\pm 1)\langle \tau_{(0)}\rangle \mp
\langle \tau_{(1)}\rangle$, $\beta=\sqrt{1+[\langle
\tau_{(2)}\rangle-14\langle
 \tau_{(1)}\rangle]/(98\langle \tau_{(0)}\rangle)}$.

These Hamiltonian solutions (\ref{12-17}) and (\ref{12-18})
  do not contain
the Lifshits-type kinetic scalar perturbations
 explaining the CMB spectrum in the Inflationary Model
 \cite{bard}; they disappear due to the
 positive energy  constraint (\ref{hg}). Therefore, the problem arises
  to reproduce the  CMB spectrum
   by  the fundamental operator quantization based on the Hamiltonian
   approach to GR.

One can construct the  Hamiltonian form of the action (\ref{1gr})
\be\label{hf} S=\int dx^0\left[\int d^3x \left(\sum_F
P_F\!\partial_0
F\!+\!C\!-\!\widetilde{N}_d\widetilde{T}^0_0\right)\!-\!P_{\varphi}\partial_0\varphi+
\frac{P_{\varphi}^2}{4\int dx^3 ({\widetilde{N}_d})^{-1}}\right],
\ee in terms of momenta $P_{ F}=[{p_{\widetilde{\psi}}},
p^i_{{(b)}},p_f]$  given by (\ref{gauge}) and $P_\vh=\partial
L/\partial (\partial_0\vh)$ ,
  where
 ${\cal C}=N^i {\widetilde{T}}^0_{i} +C_0p_{\widetilde{\psi}}+ C_{(b)}\partial_k{\bf
e}^k_{(b)}$
  is the sum of constraints
  with the Lagrangian multipliers $N^i,C_0,~C_{(b)}$ and the energy--momentum tensor
  components $\widetilde{T}^0_i$; these constraints include
    Dirac's
 constraints  (\ref{hg})  and the transversality
  $\partial_i {\bf e}^{i}_{(b)}\simeq 0$
   \cite{dir}.

 One can find
 evolution of all field variables $F(\vh,x^i)$  with respect to
 $\vh$ by the variation of the ``reduced'' action obtained as
   values of the Hamiltonian form of the initial action  (\ref{hf}) onto
 the energy constraint  (\ref{nph}):
 \be\label{2ha2} S|_{P_\vh=\pm E_\vh} =
 \int\limits_{\vh_I}^{\vh_0}d\widetilde{\vh} \left\{\int d^3x
 \left[\sum\limits_{  F}P_{  F}\partial_\vh F
 +\bar{\cal C}\mp2\sqrt{\widetilde{T}_0^0(\widetilde{\vh})}\right]\right\},
\ee
 where $\bar{\cal C}={\cal
 C}/\partial_0\widetilde{\vh}$ and $\vh_0$ is the
 present-day datum that has no relation to the  initial
 data at the beginning $\vh=\vh_I$.

\section{\label{s-4}Observational Data in  Terms of Scale-Invariant Variables}

 Let us assume that the  density
 $T_0^0=\rho_{(0)}(\vh)+T_{\rm f}$
 contains a tremendous  cosmological background
$\rho_{(0)}(\vh)$.
 The  low-energy decomposition
  of ``reduced''  action (\ref{2ha2})  $2 d\vh \sqrt{\widetilde{T}_0^0}= 2d\vh
\sqrt{\rho_{(0)}+T_{\rm f}}
 =
 d\vh
 \left[2\sqrt{\rho_{(0)}}+
 T_{\rm f}/{\sqrt{\rho_{(0)}}}\right]+...$
 over
 field density $T_{\rm f}$ gives the sum
 $S|_{P_\vh=+E_\vh}=S^{(+)}_{\rm cosmic}+S^{(+)}_{\rm
 field}+\ldots$, where the first  term of this sum
 $S^{(+)}_{\rm cosmic}= +
 2V_0\int\limits_{\vh_I}^{\vh_0}\!
 d\vh\!\sqrt{\rho_{(0)}(\vh)}$ is  the reduced  cosmological
 action,
 whereas the second one is
  the standard field action of GR and SM
 $ 
 S^{(+)}_{\rm field}=
 \int\limits_{\zeta_I}^{\zeta_0} d\zeta\int d^3x
 \left[\sum_{F}P_{ F}\partial_\eta F
 +\bar{{\cal C}}-T_{\rm f} \right]
 $ 
 in the  space with the  interval
 $
 ds^2=d\zeta^2-\sum_a[e_{(a)i}(dx^i+{\cal N}^id\zeta)]^2;
 ~\partial_ie^i_{(a)}=0,~\partial_i{\cal N}^i=0
  $
 and  conformal time
 $d\eta=d\zeta=d\vh/\rho_{(0)}^{1/2}$ as the gauge-invariant
  quantity, coordinate distance
 $r=|x|$,
 and running masses
 $m(\zeta)=a(\zeta)m_0$.
 We see that
  the  correspondence principle leads to the theory,
  where the scale-invariant conformal  variables and coordinates are
    identified  with the observable ones
    and the cosmic evolution with the evolution of masses:
  $$
\frac{E_{\rm emission}}{E_0}=\frac{m_{\rm atom}(\eta_0-r)}{m_{\rm
atom}(\eta_0)}=\frac{\vh(\eta_0-r)}{\vh_0}=a(\eta_0-r)
=\frac{1}{1+z}.
$$
The conformal observable distance  $r$ loses the factor $a$, in
comparison with the nonconformal one $R=ar$. Therefore, in this
 case, the redshift --
  coordinate-distance relation $d\eta=d\vh/\sqrt{\rho_0(\vh)}$
  corresponds to a different
  equation
  of state  in comparison with the standard one  \cite{039}.
    The best fit to the data  including
  cosmological SN observations \cite{SN2}
 requires a cosmological constant $\Omega_{\Lambda}=0.7$,
$\Omega_{\rm CDM}=0.3$ in the case of the Friedmann
``scale-variant quantities`` of standard cosmology, whereas for
the ``scale-invariant conformal
 quantities''
 these data are consistent with  the dominance of the stiff state
of a free scalar field $\Omega_{\rm Stiff}=0.85\pm 0.15$,
$\Omega_{\rm CDM}=0.15\pm 0.10$ \cite{039}. If $\Omega_{\rm
Stiff}=1$, we have the square root dependence of the scale factor
on conformal time $a(\eta)=\sqrt{1+2H_0(\eta-\eta_0)}$. Just this
time dependence of the scale factor on
 the measurable time (here -- conformal one) is used for a description of
 the primordial nucleosynthesis \cite{039}. Thus, the relative  units
 can describe all epochs including the creation of a quantum universe at
$\vh(\eta=0)=\vh_I,H(\eta=0)=H_I$ by the
  stiff state
\cite{039}. This homogeneous stiff state can be formed by a
free scalar field. 

\section{GR ``Energy'' and Creation of Universe with the Time Arrow}

 The ``reduced'' action (\ref{2ha2}) and the correspondence
 principle considered in the previous Section show us that
 the energy is the value of the scale factor canonical momentum
 \bea \label{pph}
 P_\vh&=&\frac{\partial L}{\partial (\partial_0\vh)}
 = -2V_0\partial_0\vh
 \left\langle(\widetilde{N}_d)^{-1}\right\rangle=
 -2V_0\frac{d\varphi}{d\zeta}\equiv-
2V_0 \vh'
 \eea
 obtained by the spatial averaging the energy constraint
(\ref{nph}) that takes the form
  $P^2_\vh-E_\vh^2=0$, where  $E_\vh=2\int d^3x(\widetilde{T}_0^0)^{1/2}$.
 Finally, we get the field space of events $[\vh|\widetilde{F}]$,
 where $\vh$ is the evolution parameter, and its canonical momentum
 $P_{\vh}$
 plays the role of the  Einstein-type energy.

  The primary quantization of the energy
  constraint $[\hat P^2_\vh-E_\vh^2]\Psi_L=0$ leads  to
  the unique  wave function $\Psi_L$
  of the
  collective cosmic motion.
 The secondary quantization
 $\Psi_{\rm
L}=\frac{1}{\sqrt{2E_\vh}}[A^++A^-]$ describes   creation of a
``number'' of universes
  $<0|A^+A^-|0>=N$
  from the stable Bogoliubov vacuum  $B^-|0>=0$, where $B^-$ is
   Bogoliubov's operator of annihilation of the universe
obtained by the transformation
 $ A^+=\alpha
 B^+\!+\!\beta^*B^-$ in order to diagonalize  equations of
 motion.

 This causal quantization with the minimal energy
 restricts the motion of the universe in the field space of events
 $E_\vh > 0, \vh_0>\vh_I$ and $E_\vh < 0, \vh_0<\vh_I$, and it
 leads to the arrow of the time interval $\zeta \geq 0$ as the
  quantum anomaly \cite{pvng8a}.

\section{Creation of Matter  and Initial Data of the Universe}
 The initial data $\vh_I,H_I$ of the universe can be  determined
 from the
 parameters of matter cosmologically created from the stable
 quantum
 vacuum  at the beginning of the universe.

  The Standard
 Model in the framework of the perturbation theory
 and the  operator quantization of SM
  \cite{pvn3} shows us  that  W-,Z-vector bosons
 have maximal probability of the
 cosmological creation due to their mass singularity.
 The uncertainty principle $\triangle E\cdot\triangle \eta \geq 1$
  (where $\triangle E=2M_{\rm I},\triangle \eta=1/(2H_{\rm I})$)
 testifies that these bosons can be created
 from vacuum at   the moment when   their  Compton   length
 defined by the inverse mass
 $M^{-1}_{\rm I}=(a_{\rm I} M_{\rm W})^{-1}$ is close to the
 universe horizon defined in the
 stiff state as
 $H_{\rm I}^{-1}=a^2_{\rm I} (H_{0})^{-1}$.
 Equating these quantities $M_{\rm I}=H_{\rm I}$
 one can estimate the initial data of the scale factor
 $a_{\rm I}^2=(H_0/M_{\rm W})^{2/3}=10^{-29}$ and the Hubble parameter
 $H_{\rm I}=10^{29}H_0\sim 1~{\rm mm}^{-1}\sim 3 K$ \cite{pvng7}.


 The collisions and scattering processes with the cross-section
 $\sigma \sim 1/M_{\rm I}^2$  lead to conformal
 temperature $T_c$. This temperature
   can be estimated  from the condition that  the relaxation time
 is close to the life-time of the universe, i.e.,
  from the equation in the
 kinetic theory $\eta^{-1}_{relaxation}\sim n(T_c)\times \sigma \sim H $.
 As
 the distribution functions of the longitudinal   vector bosons
 demonstrate a large contribution of relativistic momenta
 \cite{pvng7} $n(T_c)\sim T_c^3$,
 this kinetic equation  gives the temperature of
relativistic bosons $
 T_c\sim (M_{\rm I}^2H_{\rm I})^{1/3}=(M_0^2H_0)^{1/3}\sim 3 K
$ as a conserved number of cosmic evolution compatible with the SN
data \cite{039}.
 We can see that
this  value is surprisingly close to the observed temperature of
the CMB radiation
 $ T_c=T_{\rm CMB}= 2.73~{\rm K}$.
 The equations describing the longitudinal vector bosons
 in SM, in this case, are close to
 the equations that  are  used in  the
 Inflationary Model \cite{bard} for
 description of the ``power primordial spectrum'' of the CMB radiation.

  The primordial mesons before
 their decays polarize the Dirac fermion vacuum and give the
 baryon asymmetry frozen by the CP -- violation
 so that $n_b/n_\gamma \sim X_{CP} \sim 10^{-9}$,
 $\Omega_b \sim \alpha_{\rm \tiny QED}/\sin^2\theta_{\rm Weinberg}\sim
 0.03$, and $\Omega_R\sim 10^{-4}$~\cite{pvng7}.

 All these results
 testify to that all  visible matter can be a product of
 decays of primordial bosons, and the observational data on CMB
  reflect rather  parameters of the primordial bosons, than the
 matter at the time of recombination. In particular,
 the length of  the semi-circle on the surface of  the last emission of
photons at the life-time
  of W-bosons
  in terms of the length of an emitter
 (i.e.
 $M^{-1}_W(\eta_L)=(\alpha_W/2)^{1/3}(T_c)^{-1}$) is
 $\pi \cdot 2/\alpha_W$.
 It is close to the value of
 orbital momentum with the maximal $\Delta T$:
 $l_{(\Delta T_{\rm max})}\sim \pi \cdot 2/\alpha_W\sim  210 $,
 whereas $(\bigtriangleup T/T)$ is proportional to the inverse number of
emitters~
 $(\alpha_W)^3 \sim    10^{-5}$.

In relative units the temperature history of the expanding
universe looks like the
 history of evolution of masses of elementary particles in the cold
 universe with the constant conformal temperature $T_c=a(\eta)T=2.73~ {\rm K}$
 of the cosmic microwave background.

 In relative units the  nonzero shift vector and
  the scalar potentials given by Eqs. (\ref{12-17}) and (\ref{12-18})
    determine
  \cite{039} the parameter
  of spatial oscillations
  $ m^2_{(-)}=\frac{6}{7}H_0^2[\Omega_{\rm R}(z+1)^2+\frac{9}{2}\Omega_{\rm
  Mass}(z+1)]$. The redshifts in the recombination
  epoch $z_r\sim 1100$ and the clustering parameter
 $
 r_{\rm clust.}={\pi}/{ m_{(-)} }\sim {\pi}/[{
 H_0\Omega_R^{1/2} (1+z_r)}] \sim 130\, {\rm Mpc}
 $
  recently
 found in the searches of large scale periodicity in redshift
 distribution \cite{da}
 lead to a reasonable value of the radiation-type density
  (including the relativistic baryon matter one)
  $10^{-4}<\Omega_R\sim 3\cdot 10^{-3}$ at the time of this
  epoch.

\section*{Conclusions}

The
 observational astrophysical data on CMB radiation
  revealed that our Universe can be an ordinary physical
  object moving with respect to the Earth observer
   with occasional initial data.
  This revelation returns  us back to
 representations of the Poincare group as the basis of
  operator quantization that
  includes occasional gauge-invariant and
  frame-covariant initial data and their units of measurements.

 In 
 order
 to explain the World,
 a modern Laplace should ask for
{\it  the initial data of the gauge-invariant
 variables measured  in the
 relative units in the comoving reference frame of this World}.
 The statement of the problem proposes a complete separation of
 frame transformations from the gauge ones. This separation is the
 main difference of our Hamiltonian approach to GR  from
 all other ones. The result is the exact resolution of the energy
 constraint in terms of
  gauge-invariant
 variables that mean here the application of the standard theory
 of the unitary irreducible representations of the Poincare group
 based on the time-like unit vector that distinguishes the
 comoving frame in the invariant space-time where components of
 the Fock simplex are given.
 Another frame means a choice of another time-like unit vector connected with the
 first one by the Lorentz transformation that leads to the dipole
 component of the CMB temperature.

  Here we listed
  a set of numerous arguments in favor of that
 the fundamental operator quantization
  can be a real theoretical basis
 for a   further detailed investigation of astrophysical
 observational data, including
  CMB fluctuations.

\section*{Acknowledgments}

The authors are grateful to
   A.~Efremov, M. Francaviglia, Z.~Oziewicz,
 V.~Priezzhev,   and S.~Vinitsky
 for fruitful discussions. VNP thanks the Bogoliubov -- Infeld
 foundation for financial support.

\end{document}